\documentclass[sensors,article,submit,moreauthors]{Definitions/mdpi}

\firstpage{1}
\pubvolume{1}
\issuenum{1}
\articlenumber{0}
\pubyear{2026}
\copyrightyear{2026}
\datereceived{ }
\daterevised{ }
\dateaccepted{ }
\datepublished{ }

\Title{A Study of the 30 May 2026 Meteor Event in Eastern Massachusetts
Employing Data from Raspberry Boom Infrasound Sensors in the Boston Area}

\Author{Steven Africk $^{1,}$* and Abraham Loeb $^{1}$}

\AuthorNames{Steven Africk and Abraham Loeb}

\address[1]{%
$^{1}$ \quad Galileo Project, Harvard University, 60 Garden Street, Cambridge, MA 02138, USA}

\corres{Correspondence: safrick@att.net}

\abstract{Infrasound signals of one and a half cycles of a 0.4~Hz pressure pulse were detected from the 30 May 2026 meteor over Cape Cod Bay at several university and private Raspberry Boom sensors in the Greater Boston area. Our analyses of these signals demonstrate the ability to define directivity to the source of the event based on differences in the timing of the signals, to determine its identity and location by analysis of the relationship between event distance and sensor event timing, and event acoustical power by the infrasonic pressures measured at the sensor sites. The infrasound event is determined to be the atmospheric fireball blast at between 40 and 50~km in altitude with acoustic energy on the order of 5 - 14\% of the total fireball~energy.}


\let\linenumbers\relax % line numbers off
\begin{document}
\nolinenumbers

\section{Introduction}

\begin{figure}[H]
\centering
\includegraphics[width=12.0cm]{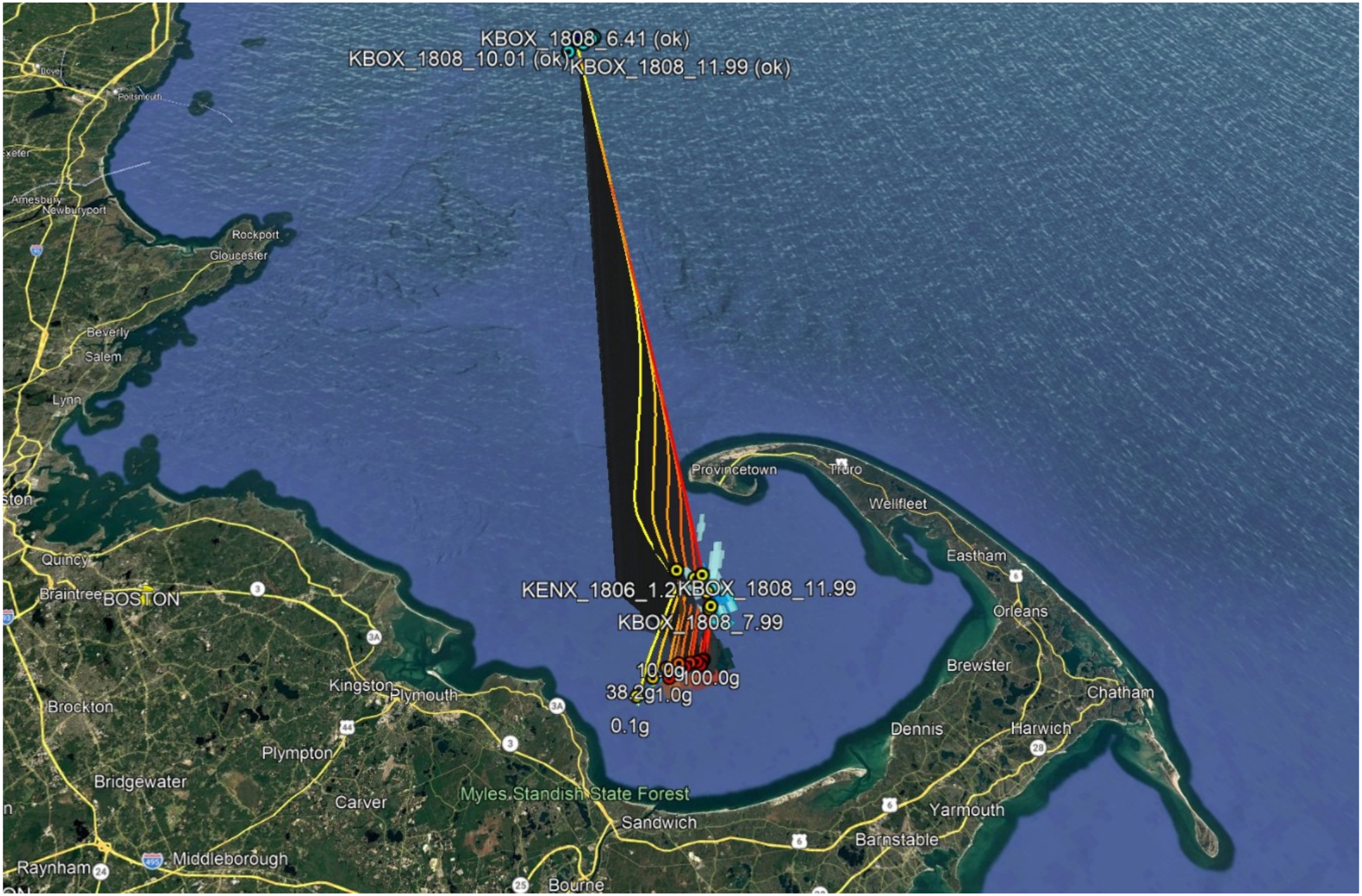}
\caption{NASA ARES reconstruction of meteor path from fireball blast to
splash.\label{fig1}}
\end{figure}

\begin{figure}[H]
\centering
\includegraphics[width=12.0cm]{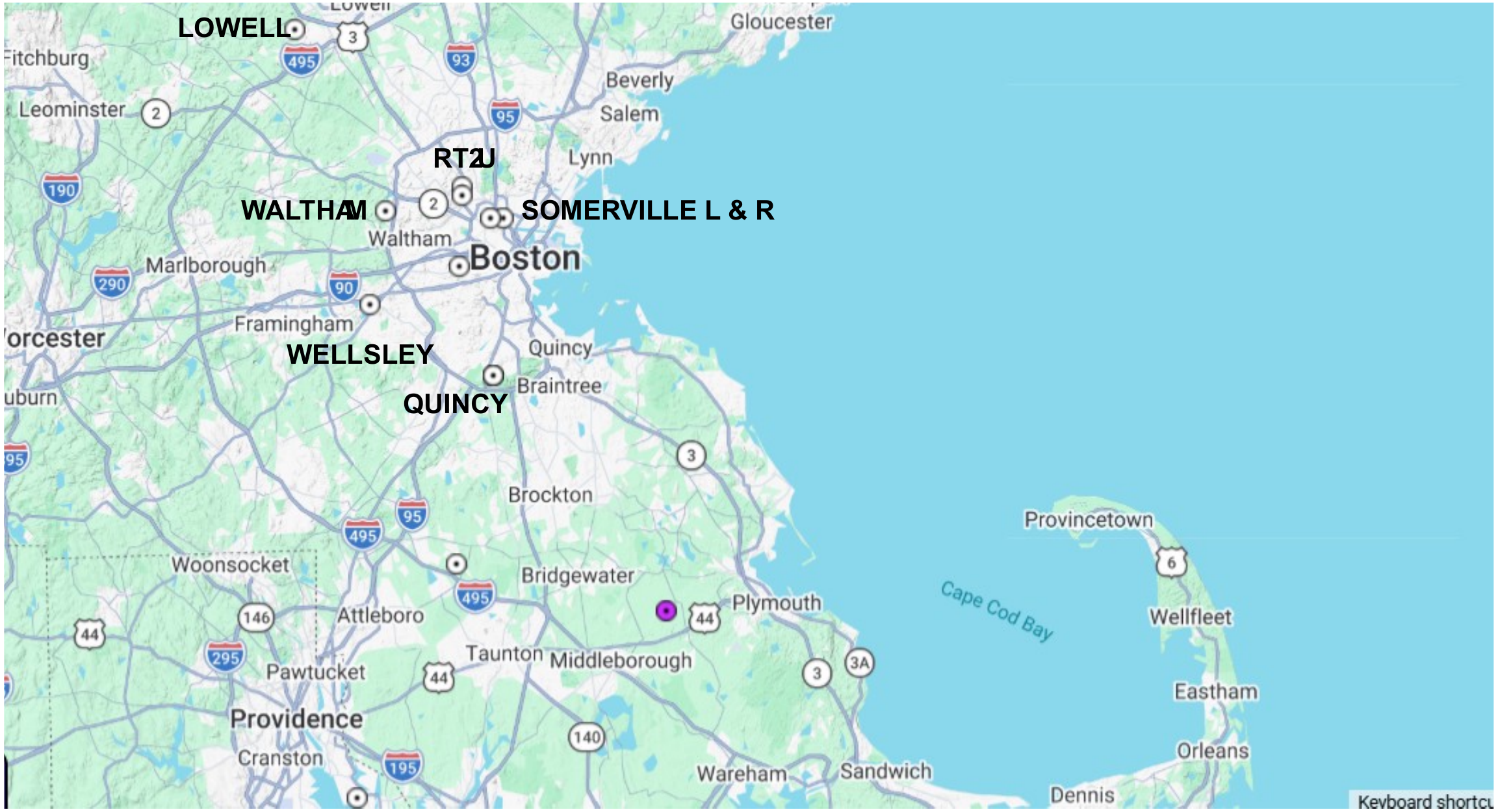}
\caption{White circles with black dots identify sites of seven Raspberry Boom sensors in
greater Boston used in this study.\label{fig2}}
\end{figure}

On 30 May 2026 a bolide (NASA ARES Event 3867 for 2026) landed in Cape Cod Bay approximately 80 km southeast of Boston, Massachusetts. The entry point in the atmosphere was originally estimated 32~km above a latitude of 42$^\circ$ and longitude of $-70.5^\circ$ but the altitude has been revised to 50~km \citep{ref1}, and the splash point was estimated by NASA to have been at latitude 41.87754$^\circ$ and longitude $-70.35239^\circ$ (Figure~\ref{fig1}). This event produced a loud impulsive sound that was  heard by hundreds of people on the East Coast at approximately 18:11 UTC which created a media sensation. It also produced a double-peaked infrasound signal approximating 1.5 cycles of a 0.4~Hz pressure pulse with amplitude ranging from about 1 to 50~Pa that was
detected by several Raspberry Boom sensors maintained by educational institutions and hobbyists in greater Boston (Figure~\ref{fig2}). These are part of a worldwide system whose
data is readily available online \citep{ref2,ref3}. Their locations can be found on the NSF-supported IRIS website \citep{ref4}. Here we report an analysis of these signals demonstrating methodologies used to study this event to characterize its location, to estimate its energy content, and to distinguish between its two possible sources, the original fireball blast in the atmosphere and the splash of fragments into Cape Cod Bay. This infrasound framework should be useful for the study of future events of this nature.

\section{Background}

\subsection{The Data}

\begin{figure}[H]
\centering
\includegraphics[width=12.0cm]{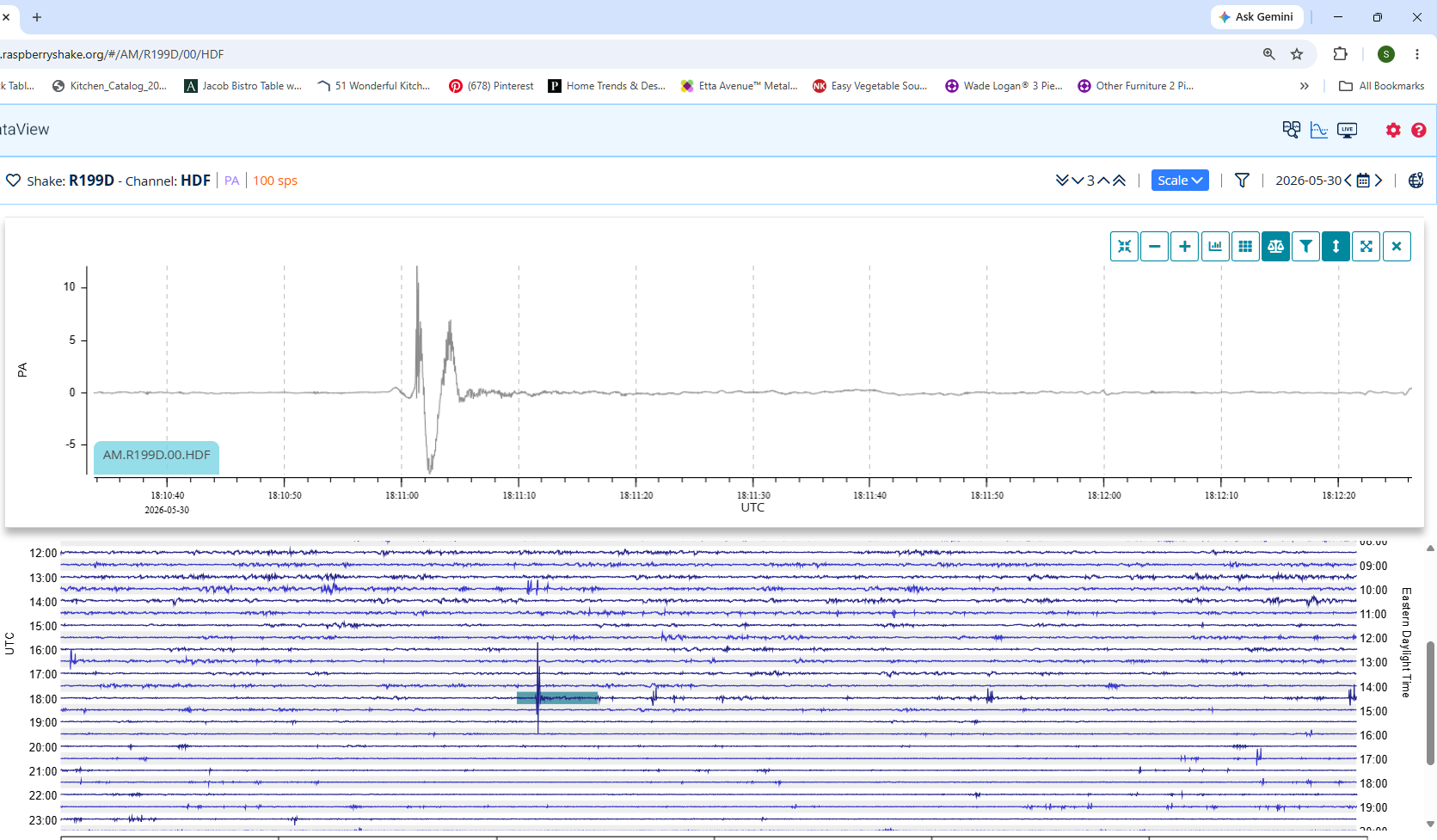}
\caption{Infrasound signal detected at SomervilleL site.\label{fig3}}
\end{figure}

\begin{figure}[H]
\centering
\includegraphics[width=9.0cm]{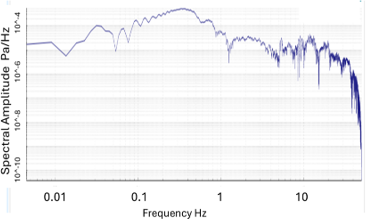}
\caption{Pressure Spectral Amplitude of signal detected at SommervilleL with peak at
0.4~Hz.\label{fig4}}
\end{figure}

A representative time series recorded at the SomervilleL sensor site is shown in
Figure~\ref{fig3}. The bottom of the figure shows the time trace of the measured pressure
for a 24-hour period with each line representing a 30-minute time window. The segment
containing the meteor signal is highlighted; it is clearly the strongest signal detected
that day. The top of the figure shows the signal in detail with a first peak at 18:11:36
UTC as measured by the position of the top of the first positive peak, some five minutes after the meteor
entered the atmosphere and exploded. It is approximately a one and half cycle of a 0.4~Hz
sine wave. A spectrum provided by the Raspberry Shake and Boom website that demonstrates a maximum at around 0.4~Hz is shown in Figure~\ref{fig4}. The peak-to-peak amplitude of this signal is some 19 Pa. Similar signals were observed at six other sites, summarized in Table~\ref{tab1}, including the time of arrival of the infrasound signal, the estimated horizonal distance to the atmospheric fireball blast and to the fragment splash site. In addition, simultaneous signals were observed in the seismic sensors included in the
Shake and Boom devices. These showed a rapid onset precisely aligned with the infrasound signals and a significantly greater frequency content suggesting that they were likely due to
mechanical excitation of the seismographs by the pressure pulse.

\begin{table}[H]
\caption{Listing of Raspberry Boom sites including timing of first and second peaks,
horizontal distance to fireball and splash points.\label{tab1}}
\begin{adjustwidth}{-\extralength}{0cm}
\newcolumntype{R}{>{\raggedleft\arraybackslash}X}
\small
\begin{tabularx}{\fulllength}{llRRRRRRR}
\toprule
\textbf{Official} & \textbf{Nickname} & \textbf{Latitude} & \textbf{Longitude}
& \textbf{First Peak} & \textbf{Second Peak} & \textbf{Distance} & \textbf{Distance}\\
\textbf{Designation} & & & & \textbf{Fractional} & \textbf{Fractional}
& \textbf{to Blast} & \textbf{to Splash}\\
 & & & & \textbf{Minutes} & \textbf{Minutes} & \textbf{(km)} & \textbf{(km)}\\
 & & & & \textbf{After 18:00} & \textbf{After 18:00} & & \\
\midrule
RF082 & Quincy      & 42.2072  & $-$71.10762 & 10.40 & 10.45 & 63.80  & 72.30\\
R0352 & Lowell      & 42.6036  & $-$71.41500 & 12.66 & 12.81 & 105.76 & 119.00\\
R9A2D & Wellesley   & 42.2882  & $-$71.29856 & 11.25 & 11.32 & 79.89  & 90.40\\
S5379 & Waltham     & 42.3964  & $-$71.27295 & 11.53 & 11.59 & 83.75  & 95.30\\
R11C3 & Rt2U        & 42.42342 & $-$71.15513 & 11.27 & 11.32 & 78.43  & 89.80\\
R199D & SomervilleL & 42.38739 & $-$71.11204 & 11.02 & 11.07 & 73.62  & 84.50\\
R5BB4 & SomervilleR & 42.38739 & $-$71.09206 & 10.98 & 11.03 & 72.54  & 83.3\\
\bottomrule
\end{tabularx}
\end{adjustwidth}
\end{table}

\section{Methods}

\subsection{Directivity to the Source}

\begin{figure}[H]
\centering
\includegraphics[width=12.0cm]{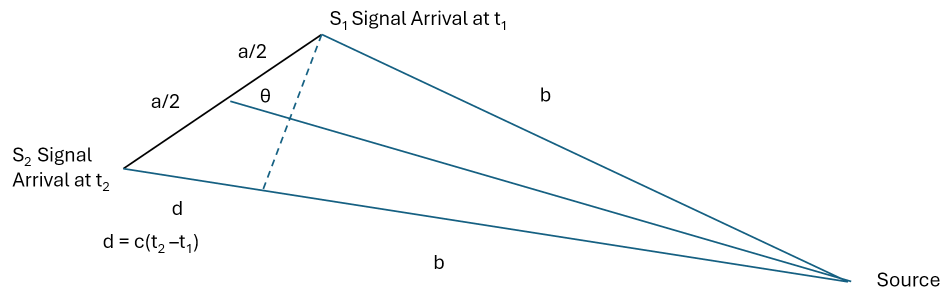}
\caption{Geometry for source directivity estimation.\label{fig5}}
\end{figure}

Data from four representative pairs of sites were used to estimate directivity to the
source based on the difference in time of arrival at the sites. The geometrical relationship
between the difference in time of arrival at two sensors and direction to a source is
sketched in Figure~\ref{fig5}, where $b$ is the distance to the source from the midpoint of
a line between two sites, $a$ is the distance between the sites, and $d = c(t_2 - t_1)$ is the path length difference, where $t_2 - t_1$ is the time difference between the signal incidence at the two sites and $c$ is the speed of sound. Using the law of cosines, the angle between the line to the source and the line between two sensors at the midpoint, $\theta$, follows:
\begin{equation}
\cos\theta = \frac{d\left(1 + d/2b\right)}
{a\sqrt{1 + \dfrac{d}{b} + \dfrac{1}{2}\left(\dfrac{d}{b}\right)^{2}
- \dfrac{1}{4}\left(\dfrac{a}{b}\right)^{2}}}\,.
\end{equation}
This angle is a function of the distance to the source $b$, which in general is not known a
priori. However, if it is known that the source is at a greater distance than the
separation between sensors and the path difference of the acoustic wave, $b \gg a$ and
$b \gg d$, we can approximate the angle simply by:
\begin{equation}
\cos\theta = \frac{d}{a}\,.
\end{equation}
In the extreme cases where the signals arrive simultaneously $d = 0$ and $\cos\theta = 0$
and $\theta = 90^\circ$ and if the signal arrives along the line separating the sensors,
$d = a$, and $\cos\theta = 1$ and $\theta = 0$. These are the only cases which are
independent of the distance to the source $b$.

This directivity is not unique. It represents a continuous possibility, rotated about the
line connecting the two sites forming a cone in three dimensions. On the surface of
the earth it represents a mirror reflection of two directional lines reflecting the positive and
negative values of the sine of the angle forming a ``V'', as both positive and negative
angles represent the same timing difference between the sites and satisfy the cosine
criteria. This ambiguity cannot be lifted by a single sensor pair measurement, but multiple
measures with components pointing at the source can lift the ambiguity as shown below.

\subsection{Event Characterization}

The timing and location of an infrasound-generating event can be characterized by a hypothesis testing analysis whereby a range of possible source positions are assumed and implied values of the source distance and timing for each are compared with the data.  This can corroborate or refine existing data and potentially determine unknown properties. As an example, we can plot known or estimated distances to the presumed event vs. time of signal arrival. For the event studied here, these curves are well approximated by straight lines with slope equal to the effective average acoustic wave speed between the event and the sensors and the time axis intercept representing zero distance and hence event time.

For example, the 30 May 2026 event has two possible origins for the infrasound, the
atmospheric fireball blast, and the subsequent splash of fragments into Cape Cod Bay.
Distance inputs to the distance--time plot from the blast is the hypotenuse measure of the
horizontal distance between the known fireball latitude and longitude and the presumed
altitudes. The effective wavespeed indicated by the slopes of the linear least-squares fits shown in the plot is a single value representing the path through the atmosphere in which, as shown in the Appendix~\ref{appA}, the wavespeed varies from about 330~m/s at 50~km, down to closer to 300~m/s at 20~km and back up to 340~m/s at sea level so that we should expect to find an effective average wavespeed on the order of 320~m/s. It should be noted that the path through the atmosphere can be affected by refraction, or the bending of the acoustic rays from a straight line to areas of lower soundspeed, so that the actual path can be longer than the abovementioned straight-line calculations. This suggests that these wavespeed estimates, relying on shorter path lengths, can underestimate the effective wave speed. For present purposes we shall assume the differences from linearity are small and not include them in the analysis.

The splash scenario does not have this complication as it is an entirely sea level event,
but it contains another very significant unknown, the time of flight of meteor fragments
during dark flight from the fireball to the splash. This problem is exacerbated by the fact
that the flight times are a function of fragment mass and that the number of larger
fragments that were more likely to generate a strong infrasound signal is not well known. A
crude estimate of this timing assuming fireball blast at 50~km altitude, a straight-line
descent, and order of magnitude fragment speeds of Mach~1 and Mach~2 yields 1.3--2.6
minutes. Consequently, we should expect a fit with event time for the splash hypothesis on
the order of one or two minutes after the fireball. A third possibility, a sonic boom, can be ruled out as the source of infrasound, although one was heard in the audio range. It would have generated a pressure pulse that trails the supersonic fragments, which would have generated signals traveling roughly north to south.  However, as shown below, the arrival times are consistent with a single event south of the Boston area.

\subsection{Source Acoustic Energy Estimate}

To calculate the acoustic energy of the fireball blast, we assume the infrasonic pulse has
acoustic energy $E_{\mathrm{acoustic}}$  distributed into a solid angle $\Omega$, and lasts for
time $\tau$. Then the average acoustic intensity at range $R$ is
\begin{equation}
I \approx \frac{E_{\mathrm{acoustic}}}{\Omega R^{2} \tau}\,.
\end{equation}
For a plane acoustic wave,
\begin{equation}
I = \frac{p_{\mathrm{rms}}^{2}}{\rho_{\mathrm{air}} c_{\mathrm{air}}}\,,
\end{equation}
so that
\begin{equation}
p_{\mathrm{peak}} = \sqrt{\frac{2 \rho_{\mathrm{air}} c_{\mathrm{air}}
E_{\mathrm{acoustic}}}{\Omega R^{2} \tau}}\,,
\end{equation}
and
\begin{equation}
E_{\mathrm{acoustic}} = \frac{\Omega R^{2} \tau}
{2 \rho_{\mathrm{air}} c_{\mathrm{air}}}\, p_{\mathrm{peak}}^{2}\,,
\end{equation}
where
\begin{itemize}[leftmargin=*,topsep=3pt,itemsep=1pt]
\item $p_{\mathrm{peak}} = \sqrt{2}\, p_{\mathrm{rms}}$ Pa;
\item $\rho_{\mathrm{air}} c_{\mathrm{air}} = 410$~Pa\,s/m is the product of the air mass
density and the sound speed;
\item $R = 80$~km, a value typical of distance from the fireball blast and the splash to
the sensors;
\item $\Omega = 4\pi$ is the solid angle into which the blast energy is distributed; For a splash, the appropriate value is $2\pi$ 
\item $\tau = 4$~s representative of the one and a half cycles of the observed signals;
\item $E_{\mathrm{acoustic}}$ is acoustic energy generated by the event, \end{itemize}
so that the acoustic energy is estimated by
\begin{equation}
E_{\mathrm{acoustic}} = 3.92 \times 10^{8}\, p_{\mathrm{peak}}^{2}\,.
\label{eq7}
\end{equation}

\section{Results}

\subsection{Directivity to Source}

\begin{figure}[H]
\centering
\includegraphics[width=12.0cm]{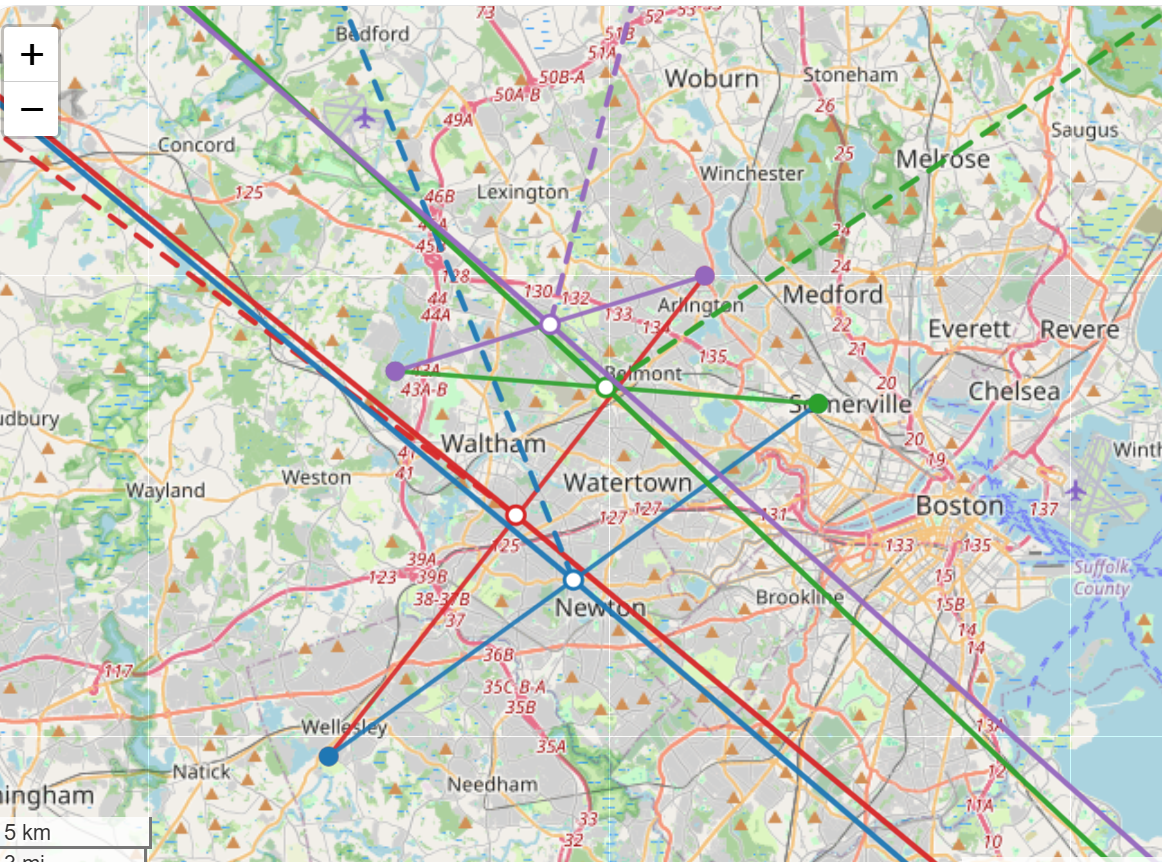}
\caption{Sensor pair locations and positive (solid) and negative (dotted) direction lines
illustrating the mirrorlike ambiguity in the directivity calculation.\label{fig6}}
\end{figure}

\begin{figure}[H]
\centering
\includegraphics[width=12.0cm]{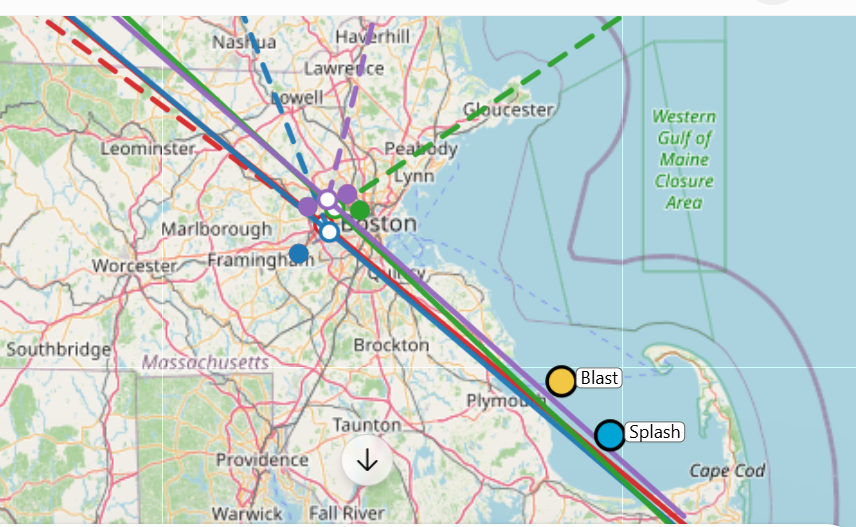}
\caption{Positive angle direction lines converging on the direction of the meteor sites
with negative angle lines diverging away from the bay.\label{fig7}}
\end{figure}

Table~\ref{tab2} lists four representative station pairs and the angles calculated by
Equation~(2) used to generate direction lines projected to the source. Figure~\ref{fig6} shows the four pairs with a line drawn between the sensors and the two direction lines from the
midpoint, with the solid lines and dotted lines representing the positive and negative
angles. Figure~\ref{fig7} shows the two sets of direction lines on a larger geographical
scale including Cape Cod Bay and the sites for the fireball and splash. Here, the
ambiguities are lifted as the positive-angle lines converge and point toward the fireball
and splash points while the negative angle (dotted) lines diverge in directions away from
the bay.

\begin{figure}[H]
\centering
\includegraphics[width=10.0cm]{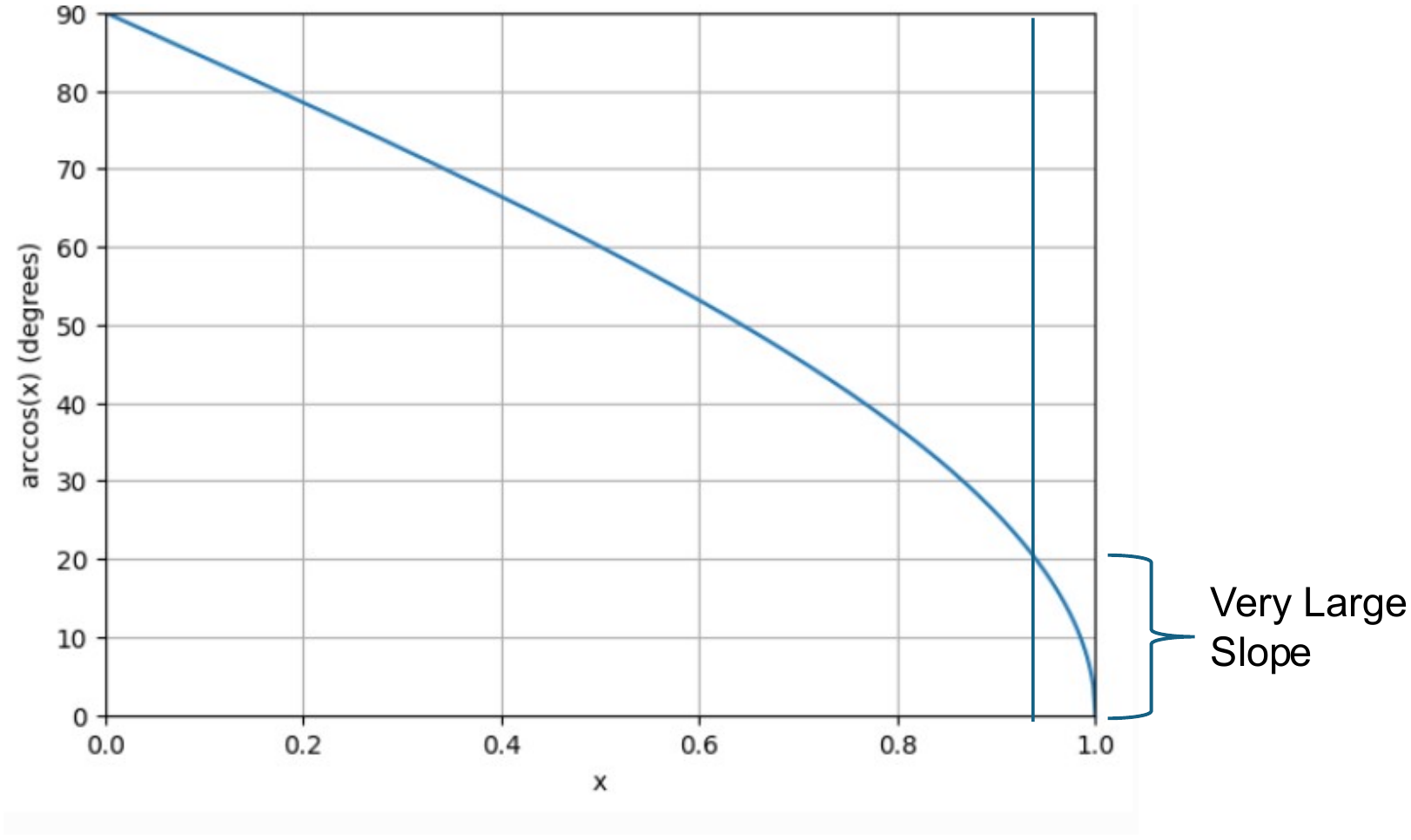}
\caption{Inverse cosine function showing large slope below
$20^\circ$.\label{fig8}}
\end{figure}

The set of positive-angle directional lines all point in the direction of the two possible
sources. However, as only two cross at a single point---between the possible
sources---this is a failure of triangulation as it does not provide a clear discrimination between the two possible sources. This is due to the inherently approximate nature of the method, the closeness of the sensor sites, and the near alignment of the sensor group with the two possible sources. This method, while able to point to a source, cannot by itself distinguish between the two sites in this case. The addition of a sensor pair further down the coast would almost certainly cross these lines and generate true triangulation. As an aside, in the process of these calculations a significant limitation of this method was encountered. As shown in Figure~\ref{fig8}, the slope of the inverse cosine function is very large below about $20^\circ$, making it extremely sensitive to small variations in the input
$d/a$ in this range. This occurs when the line between the two sensors is close to
collinear with the line to the source. Pairs with this property were not included in this analysis.

\begin{table}[H]
\caption{Sensor pairs and their angles to event.\label{tab2}}
\begin{adjustwidth}{-\extralength}{0cm}
\scriptsize
\centering
\begin{tabular}{@{}lccc@{\hspace{8pt}}lccc@{\hspace{8pt}}rrrrrrr@{}}
\toprule
\multicolumn{4}{c}{\textbf{Site 1}} & \multicolumn{4}{c}{\textbf{Site 2}}
& \multicolumn{2}{c}{\textbf{Time Difference}}
& \boldmath$d=c\,\Delta t$ & \boldmath$a$ & \boldmath$\cos|\theta|$
& \multicolumn{2}{c}{\boldmath$|\theta|$}\\
\cmidrule(r){1-4}\cmidrule(r){5-8}\cmidrule(r){9-10}\cmidrule(l){14-15}
\textbf{Name} & \textbf{Time} & \textbf{Lat.} & \textbf{Long.}
& \textbf{Name} & \textbf{Time} & \textbf{Lat.} & \textbf{Long.}
& \textbf{(min)} & \textbf{(s)} & \textbf{(km)} & \textbf{(km)} &
& \textbf{(rad)} & \textbf{(deg)}\\
\midrule
Wellesley & 11.25  & 42.2882 & $-$71.29856 & Rt2U        & 11.271 & 42.42342 & $-$71.1551
 & $-$0.02 & $-$1.26 & $-$0.43 & 19.10 & $-$0.02 & 1.59 & 91.29\\
Wellesley & 11.25  & 42.2882 & $-$71.29856 & SomervilleL & 11.025 & 42.38739 & $-$71.112
 & 0.23 & 13.50 & 4.63 & 18.89 & 0.25 & 1.32 & 75.80\\
Waltham   & 11.528 & 42.3964 & $-$71.27295 & SomervilleL & 11.025 & 42.38739 & $-$71.112
 & 0.50 & 30.18 & 10.35 & 13.25 & 0.78 & 0.67 & 38.62\\
Waltham   & 11.528 & 42.3964 & $-$71.27295 & Rt2U        & 11.271 & 42.42342 & $-$71.1551
 & 0.26 & 15.42 & 5.29 & 10.13 & 0.52 & 1.02 & 58.52\\
\bottomrule
\end{tabular}
\end{adjustwidth}
\noindent{\footnotesize Signal times are fractional minutes after 18:00 UTC.}
\end{table}

\subsection{Event Characterization}

\begin{figure}[H]
\centering
\includegraphics[width=12.5cm]{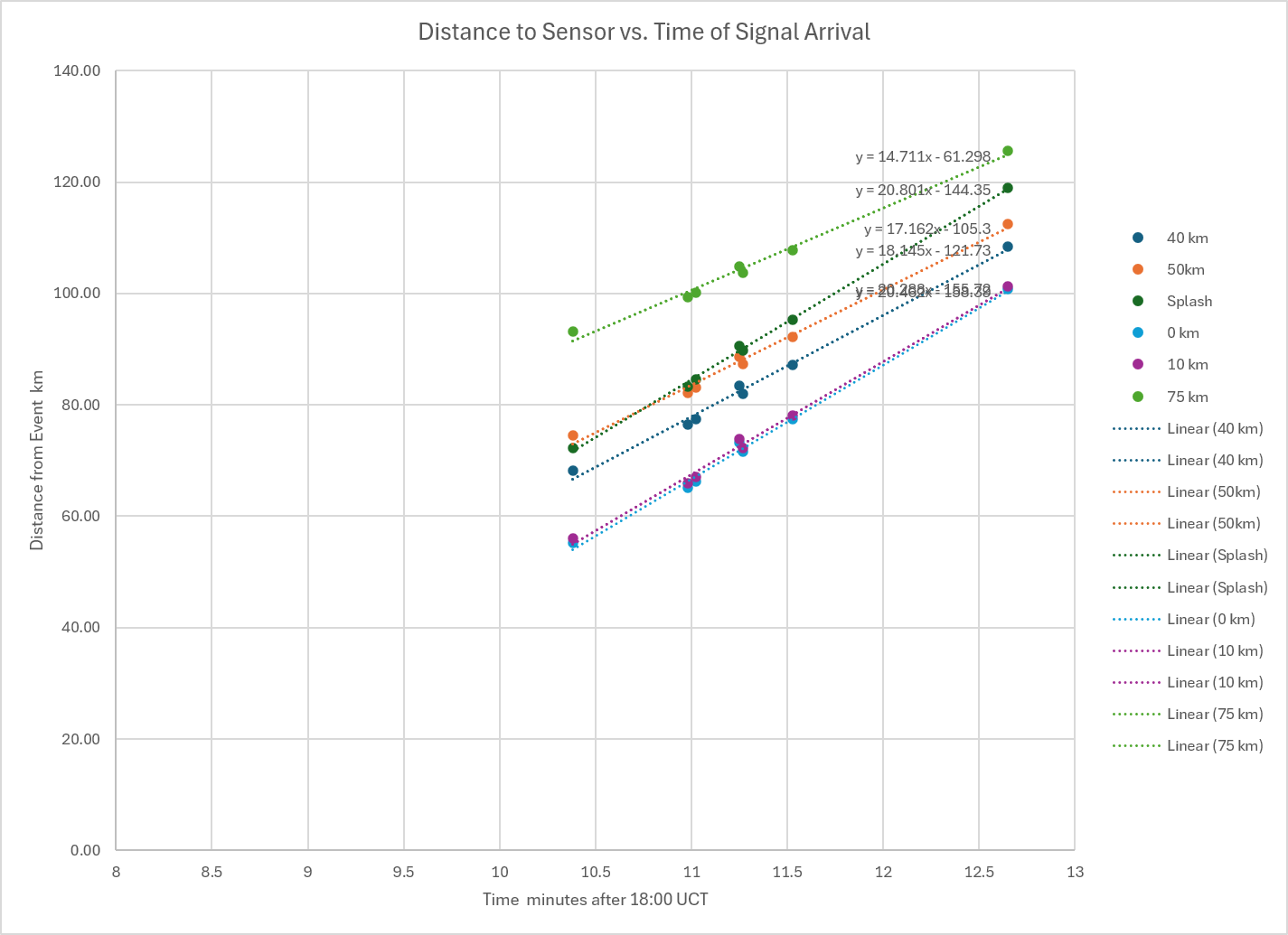}
\caption{Distance vs. time for six source scenarios.\label{fig9}}
\end{figure}

As an example of the distance--time characterization method, we consider several candidate scenarios. Figure~\ref{fig9} shows the distance--time relationships for six scenarios,
fireball blast initiated at 0, 10, 40, 50 and 75~km altitude, and splash at NASA sea level
coordinates. The linear average wavespeed determined by the fitted slope and event timing by the time axis intercept are summarized in Table~\ref{tab3}. The best fits to the data are for 40 and 50~km altitude. For the former, the implied wavespeed of 302~m/s is some 5\% below the 318~m/s estimated in the Appendix~\ref{appA}. The implied fireball blast time of 6.7 minutes after 18:00 UTC is somewhat greater than the NASA time estimate of 6 minutes based on optical data, and some tenths of a minute later by other estimates. The fit for 50~km altitude predicts a wavespeed of 286~m/s, slightly lower (by 11\%) than the average wavespeed for this path of 318~m/s estimated in Appendix~\ref{appA}. The implied event time of 6.1 minutes after 18:00 UTC is more consistent with the optical data \citep{ref1}. Fits for 0 and 75~km altitude predict times of 7.7 and 4.2 minutes, which are clearly incorrect, ruling out these scenarios. Thus, this analysis suggests that the altitude of the fireball blast was in the range of 40 to 50~km, consistent with NASA's revised value.

The fit for the splash scenario is also very good. The implied wavespeed of 346~m/s is very
close to the speed of sound at sea level which is appropriate for this scenario, and the
implied time of 6.9 minutes is 0.9 minutes after the nominal fireball blast time which is
consistent with estimates of the time of flight of the fragments to the splash site on the
order of minutes, depending on the fragment mass \citep{ref1}. That this fit is similar to
the best fireball blast fits may in part be due to the similarities of the presumed
distances, as while the splash point is farther from the sensors along the surface, the
inclusion of the altitudes brings the distances closer together. As a consequence of this
good fit, the splash hypothesis cannot be ruled out on the basis of distance--time analysis
in this case.

\begin{table}[H]
\caption{Implied acoustic wavespeeds and event timing for six event
scenarios.\label{tab3}}
\newcolumntype{C}{>{\centering\arraybackslash}X}
\begin{tabularx}{\textwidth}{lCC}
\toprule
\textbf{Event Candidates} & \textbf{Implied Acoustic Wave Speed (m/s)}
& \textbf{Implied Time, Minutes After 18:00 UTC}\\
\midrule
Blast at 0 km altitude  & 341 & 7.7\\
Blast at 10 km altitude & 338 & 7.7\\
Blast at 40 km altitude & 302 & 6.7\\
Blast at 50 km altitude & 286 & 6.1\\
Blast at 75 km altitude & 245 & 4.2\\
Splash                  & 346 & 6.9\\
\bottomrule
\end{tabularx}
\end{table}

\subsection{Source Acoustic Energy Estimate}

The fraction of the blast energy converted to the infrasound pulse can be estimated based
on the measured pressures, summarized on Table~\ref{tab4}. These vary considerably and in a manner not related to the different distances to the two site candidates, most likely due to
differences in sensitivities among site sensors. For example, there may be differences in the sensor calibrations and/or their placement, as those outdoors with good access to
the atmosphere will measure greater, and possibly more accurate, pressures than, e.g., those inside buildings. This suggests that the larger measured pressures are more representative although possible signal amplification by constructive interference of multiple paths cannot be ruled out. The largest three values range from about 50 to 90~Pa peak-to-peak and we can take 70~Pa peak-to-peak, or 35~Pa peak as an upper estimate. The average peak value of about 20~Pa may be more representative of all sensors. Finally, while the distances from the sensor sites and the two candidate sources differ, for this order of magnitude calculation we will use 80~km as a representative value.

% NOTE: Table 4 is an EMPTY table stub in Meteor_Paper9-11-2026_AviChangesInc.docx.
% The values below are derived from the peak-to-peak column of Table 1 (peak = half of
% peak-to-peak), which reproduces the numbers quoted in the text above. PLEASE VERIFY.
\begin{table}[H]
\caption{Peak-to-peak and peak pressures measured at seven sensors.\label{tab4}}
\newcolumntype{C}{>{\centering\arraybackslash}X}
\begin{tabularx}{\textwidth}{lCC}
\toprule
\textbf{Site} & \textbf{Peak-to-Peak Pressure (Pa)} & \textbf{Peak Pressure (Pa)}\\
\midrule
Quincy      & ---  & ---\\
SomervilleR & 53.6 & 26.8\\
SomervilleL & 19.3 & 9.7\\
Wellesley   & 41.3 & 20.7\\
Rt2U        & 89.2 & 44.6\\
Waltham     & 45.8 & 22.9\\
Lowell      & 3.6  & 1.8\\
\bottomrule
\end{tabularx}
\end{table}

Using the representative peak pressure of $p_{\mathrm{peak}} = 35$~Pa by Equation~(\ref{eq7}) we find $E_{\mathrm{acoustic}} = 4.8 \times 10^{11}$~J, which is 10\% of the total blast energy, estimated by CNEOS at $4.6 \times 10^{12}$~J \citep{ref5}. This fraction is slightly above the high end of the expected range based on historical bolide data of 1.7--8\% \citep{ref6}. Taking instead the average peak pressure values, closer to 20~Pa peak, this estimate yields $1.7 \times 10^{11}$~J, or 3.7\% of the total blast energy which is well within the expected range.

These values rule out the splash scenario, as they are orders of magnitude greater than the
kinetic energy the fragments could have contained as they splashed down. The total mass of fragments, about 500~kg \citep{ref1}, if contained in a single object, would have required a splash speed of about 26,000~m/s to have energy equal to the low estimate of the \textit{acoustic} energy. This is greater than the lower limit of atmosphere entry speed estimated by NASA \citep{ref1}. But due to the inefficiency of the conversion of total to acoustic energy, in order to produce this much acoustic energy the splash would have had to contain an order of magnitude greater total energy, requiring roughly an additional factor of three in speed. While the fragments obviously were traveling at supersonic speeds to have generated the sonic booms that were heard, given the reality of atmospheric drag it is extremely unlikely that their speed at splash was sufficient to have generated the infrasonic acoustic energy. 

Given this conclusion, the energy calculation requires a refinement to take into account the estimated representative distance to the fireball blast at altitude of 50~km, which is some 93~km, or about 16\% greater than the estimate of 80~km used initially. This increases the predicted acoustic energy by 35\%, so that the higher value becomes $6.5 \times 10^{11}$~J or 14\% of
the total energy. The lower estimate is increased to $2.3 \times 10^{11}$~J, or 5\% of the
total energy, still within the expected range.

\section{Discussion}

The directivity to source calculations generate direction lines that point in the direction
of the both the fireball and the splash but due to their almost collinearity with the
direction lines, the relatively close spacing sensors, and the approximate nature of the
algorithm, could not triangulate and distinguish between them. However, had there be one or more sensor pairs sufficiently south of Boston metro that generated direction lines in a
Northeastern direction, these would have crossed the existing ones and provided
triangulation and distance to the source. Thus, while this analysis did not demonstrate
triangulation, it strongly suggests that it should be possible with more geographically
dispersed infrasound sensors. Nevertheless, the serendipitous density of sensors in the
Boston area allowing the analyses made here may not be possible elsewhere where there are fewer sensors. On the other hand, the long-range propagation of infrasound may allow these analyses for more sparse sensors at greater distances to the event.

Distance--time analysis for the fireball shows good compatibility with known and estimated wavespeeds and event time for altitude between 40 and 50~km, while fits to scenarios above and below this range are poor and are thereby ruled out. This agrees with the estimated altitude. This demonstrates the potential for it to be a useful tool in either corroborating other descriptions of similar events, to refine otherwise known parameters, or to measure unknown parameters.

Energy analysis based on amplitudes of the infrasonic signals provides an independent
order-of-magnitude estimate of the acoustic energy on the order of
$2.3$--$6.5 \times 10^{11}$~J, or 5--14\% the blast energy. These estimates are at the high
end of the range of expected conversion efficiencies \citep{ref6}. This energy estimate is far greater than that of the kinetic energy of the fragments in dark flight and essentially eliminates the possibility that the splash was the source of the infrasound.

\section{Conclusions}

These analyses demonstrate that detections of infrasound signals from the meteor event of
30 May 2026 by several university and hobbyist Raspberry Boom sensors in the Boston area
provide significant information about the event. Three capabilities of the analysis are demonstrated. First, signal arrival times at pairs of sensors can establish accurate direction to the source. And while triangulation was not achieved, the analysis suggests that it should be possible with more diverse sensor geometries. Second, the relationship between distance to source and signal arrival times can provide useful information about the event's identity, location, and timing. Finally, signal pressure amplitudes can provide estimates of event acoustic energy. That these sensors are geographically distributed globally and that their data are readily available online suggests that at least some of the methods developed here can be applied elsewhere, including characterization of infrasound signals of unknown origin.

\vspace{6pt}

%=================================================================
% Back matter
\authorcontributions{Conceptualization, S.A.; methodology, S.A; software, S.A; formal analysis, S.A; resources, S.A; data curation, S.A; writing---original draft preparation, S.A; writing---review and editing, S.A; supervision, S.A; project administration, S.A. and A.L.; funding acquisition, A.L. All authors have read and agreed to the published version of the manuscript.}

\funding{This work was supported by the Galileo Project at Harvard University.}

\dataavailability{The infrasound waveforms analyzed in this study are publicly available
from the Raspberry Shake data view service \citep{ref2,ref3} and the IRIS station map
\citep{ref4}. Event parameters are available from NASA ARES \citep{ref1} and CNEOS
\citep{ref5}.}

\conflictsofinterest{The authors declare no conflicts of interest.}

%=================================================================
\appendixtitles{yes}
\appendixstart
\appendix
\section[\appendixname~\thesection]{Average Speed of Sound from 50 km to Sea Level}
\label{appA}

\begin{figure}[H]
\centering
\includegraphics[width=10.0cm]{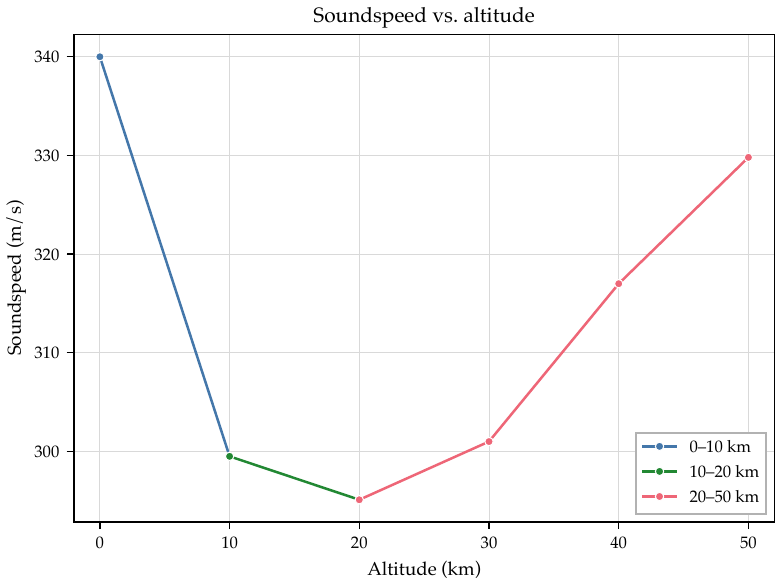}
\caption{Simplified speed of sound profile with altitude.\label{figA1}}
\end{figure}

To estimate the average speed of sound from 50~km to sea level, consider the profile vs.
altitude summarized in Figure~\ref{figA1} \citep{ref7}. For simplicity we assume that it
varies linearly with altitude and divide the range from 50 to 0~km into three straight line
segments.

Modeling the soundspeed in each segment as a linear function $c_i(x) = m_i x + b_i$ we find
the values in Table~\ref{tabA1}.

\begin{table}[H]
\caption{Parameters of three segments of soundspeed vs. altitude.\label{tabA1}}
\newcolumntype{C}{>{\centering\arraybackslash}X}
\begin{tabularx}{\textwidth}{lCC}
\toprule
\textbf{Segment (km)} & \textbf{Slope (10\boldmath$^{-3}$ s\boldmath$^{-1}$)}
& \textbf{Endpoint Wavespeed (m/s)}\\
\midrule
0--10  & $-$4.05 & 340.3, 299.5\\
10--20 & $-$0.44 & 299.5, 295.1\\
20--50 & 1.201   & 295.1, 329.8\\
\bottomrule
\end{tabularx}
\end{table}

Integrating to find the time of flight
\begin{equation}
dt = \frac{dx}{c(x)}\,,
\end{equation}
we find for each segment between $(x_1,t_1)$ and $(x_2,t_2)$ with slope $m_{1,2}$ the slope
between these points:
\begin{equation}
\int_{t_1}^{t_2} dt = t_2 - t_1
= \int_{x_1}^{x_2} \frac{dx}{m_{1,2}\,x + b_{1,2}}
= \frac{1}{m_{1,2}} \ln\!\left(\frac{c_2}{c_1}\right)\,.
\end{equation}
Adding the three times yields 157~seconds, so that for a wave propagating over 50~km yields
and average wavespeed of 318~m/s.

%=================================================================
\begin{adjustwidth}{-\extralength}{0cm}

\reftitle{References}

\PublishersNote{}
\end{adjustwidth}


\begin{thebibliography}{999}

\bibitem{ref1}
NASA Astromaterials Research and Exploration Science (ARES), Meteorite Falls: Cape Cod Bay,
MA. Available online: \url{https://ares.jsc.nasa.gov/meteorite-falls/events/cape-cod-bay-ma}
(accessed on 11 September 2026).

\bibitem{ref2}
Raspberry Shake Data View. Available online: \url{https://dataview.raspberryshake.org/}
(accessed on 11 September 2026).

\bibitem{ref3}
Raspberry Shake Manual, Web Services: Station and Earthquake Map (StationView). Available
online: \url{https://manual.raspberryshake.org/webServices.html} (accessed on 11 September
2026).

\bibitem{ref4}
IRIS/EarthScope GMAP Station Map. Available online: \url{https://ds.iris.edu/gmap/}
(accessed on 11 September 2026).

\bibitem{ref5}
CNEOS Fireballs Database, Jet Propulsion Laboratory. Available online:
\url{https://cneos.jpl.nasa.gov/fireballs/} (accessed on 11 September 2026).

\bibitem{ref6}
Anghel, S.; Drolshagen, E.; Ott, T.; Birlan, M.; Colas, F.; Nedelcu, D.A.; Koschny, D.;
Zanda, B.; Bouley, S.; Jeanne, S.; et~al. Energy signature of ton TNT-class impacts:
Analysis of the 2018 December 22 fireball over Western Pyrenees. \textit{Mon. Not. R.
Astron. Soc.} \textbf{2021}, \textit{508}, 5716--5733.

\bibitem{ref7}
U.S. Standard Atmosphere, 1976; U.S. Committee on Extension to the Standard Atmosphere

(COESA), NOAA/NASA/U.S. Air Force; U.S. Government Printing Office: Washington, DC, USA,
1976.

\end{thebibliography}
\end{document}